\documentclass{article}

\global\arraycolsep=1pt
\usepackage{amsmath}
\usepackage{amssymb}
\usepackage{graphicx}

\newcommand{\beq}{\begin{eqnarray}}
\newcommand{\eeq}{\end{eqnarray}}
\renewcommand{\tilde}{\widetilde}

\begin{document}

\begin{titlepage}
\begin{flushright}
\normalsize
PUPT-2424
\end{flushright}
\vfil

\bigskip

\begin{center}
\LARGE Exact ABJM Partition Function from TBA
\end{center}

\vfil
\medskip
\begin{center}
\def\thefootnote{\fnsymbol{footnote}}

Pavel Putrov$^\clubsuit$ and Masahito Yamazaki$^\spadesuit$
\end{center}

\medskip
\begin{center}
\small
$^\clubsuit$ 
D\'{e}partement de Physique Th\'{e}orique et Section de Math\'{e}matiques,
\\ Universit\'{e} de Gen\`{e}ve, Gen\`{e}ve,  CH-1211, Switzerland

$^\spadesuit$
Princeton Center for Theoretical Science, Princeton University, NJ 08544, USA
\end{center}

\bigskip

\thispagestyle{empty}
\setcounter{tocdepth}{1}

\bigskip

\begin{center}
{\bfseries Abstract}
\end{center}

\bigskip

We report on the exact computation of the $S^3$ partition function
of $U(N)_{k}\times U(N)_{-k}$ ABJM theory for $k=1, N=1, \ldots, 19$.
The result is a polynomial in $\pi^{-1}$ with rational coefficients.
As an application of our results we numerically determine the
coefficient of the membrane 1-instanton correction to the partition function.
\vspace{3cm}

\end{titlepage}

\newpage
\setcounter{page}{1}

\section{Introduction and Summary}

It has recently been discovered that the partition function of a 
Chern--Simons--matter (CSM) theory with
$\mathcal{N}\ge 2$ supersymmetry on a three-dimensional sphere reduces
to a matrix integral \cite{kwy,jafferis,hama}. These matrix integrals
are powerful quantitative tools to analyze CSM theories, 
and 
has lead to a number of important results, including the successful derivation of the $N^{3/2}$ behavior \cite{kt} and various precise checks of the AdS$_4$/CFT$_3$ correspondence (see \cite{mp,dmp,hklebanov} and subsequent works).

In this paper we study the CSM theory with the highest amount of
supersymmetry ($\mathcal{N}\geq 6$), namely the ABJM theory \cite{abjm}.
It has gauge group $U(N)_k\times U(N)_{-k}$, where $k$ is the level of
the Chern-Simons term.
Since ABJM theory is the worldvolume theory of multiple M2-branes,
it is natural to ask if we could extract any useful data
about M-theory from the three-sphere partition function of the ABJM theory.


In M-theory we have non-perturbative corrections from
membrane instantons. This is reflected in the three-sphere partition
function 
as an expansion of the terms of order  $e^{-\sqrt{N/k}}$ \cite{Drukker:2011zy}. 
However, this expansion is not directly captured in most of the previous analysis of the 
three-sphere partition function, where we take the t' Hooft limit 
$N, k$ large with $N/k$ kept finite.
Instead we need to take the M-theory limit, with $N$ large and $k$ kept
finite. The leading $N$ contribution in this limit is determined by \cite{hklebanov}
and the all order $1/N$ expansion in \cite{Marino:2011eh}.
Moreover the paper \cite{Marino:2011eh} discuss the 
non-perturtbative instanton correction in an expansion around $k=0$.
However the for the most interesting case of $k$ finite, the general results
on the non-perturbative corrections are still lacking.
To answer this question
it will be of great help to systematically compute 
the behavior of the three-sphere partition function 
for finite $N$ and $k$.


In this brief note we report on the exact computation of 
$S^3$ partition function $Z(N)$ of the $k=1$ ($\mathcal{N}=8$) ABJM theory for $N=1, \ldots, 19$, based on the 
Fermi gas approach of \cite{Marino:2011eh} and the 
TBA-like equations of \cite{Zamolodchikov:1994uw,Tracy:1995ax}.\footnote{See \cite{okuyama} for exact computation for $N=2$ and general $k$, and \cite{hanada} for numerical calculations.}

Our results are given as follows:
{\small
\begin{align}\label{main}
\begin{aligned}
 Z(1)&=\cfrac{1}{4} \ ,\qquad 
 Z(2)=\cfrac{1}{16 \pi } \ ,\qquad 
 Z(3)=\cfrac{-3+\pi }{64 \pi } \ , \qquad
 Z(4)=\cfrac{-{\pi ^2}+{10}}{1024{\pi ^2}} \ , \qquad
 Z(5)=\cfrac{26+20 \pi -9 \pi ^2}{4096 \pi ^2}, \\
 Z(6)&=\cfrac{78-121 \pi ^2+36 \pi ^3}{147456 \pi ^3} \ , \qquad
 Z(7)=\frac{-126+174 \pi +193 \pi ^2-75 \pi ^3}{196608 \pi ^3} \ , \\
 Z(8)&=\frac{876-4148 \pi ^2-2016 \pi ^3+1053 \pi ^4}{18874368 \pi ^4} \ , \qquad
 Z(9)=\frac{4140+8880 \pi -15348 \pi ^2-13480 \pi ^3+5517 \pi ^4}{75497472 \pi ^4} \ , \\
 Z(10)&=\frac{16860-136700 \pi ^2+190800 \pi ^3+207413 \pi ^4-81000 \pi ^5}{7549747200 \pi ^5} \ , \\
 Z(11)&=\frac{-122580+381900 \pi +837300 \pi ^2-1289300 \pi ^3-1091439 \pi ^4+447525 \pi ^5}{30198988800 \pi ^5}\ , \\
 Z(12)&=\frac{626760-8856300 \pi ^2-18446400 \pi ^3+35287138 \pi ^4+30204000 \pi ^5-12504375 \pi ^6}{4348654387200 \pi ^6}\ , \\
 Z(13)&=\frac{1563480+6714000 \pi -17252100 \pi ^2-40746000 \pi ^3+49141894 \pi ^4+45780780 \pi ^5-18083925 \pi ^6}{5798205849600 \pi ^6} \ , \\
  Z(14)&=(21382200-421152060 \pi ^2+1918350000 \pi ^3+2614227910 \pi ^4-\\ &\left. -5654854800 \pi ^5-3965159223 \pi ^6+1732468500 \pi ^7)\right/(3409345039564800 \pi ^7) \ , \\
   Z(15)&=(-222059880+1271579400 \pi +3613033620 \pi ^2-12266517900 \pi ^3-17757814914 \pi ^4+\\
 &\left.+28941378130 \pi ^5+21727092861 \pi ^6-9162734175 \pi ^7)\right/(13637380158259200 \pi ^7) \ , \\
 Z(16)&=(288454320-8196414240 \pi ^2-54540622080 \pi ^3+83379537976 \pi ^4+337956998400 \pi ^5-310977507352 \pi ^6-\\
 &\left.-354450849984 \pi ^7+132764935275 \pi ^8)\right/(872792330128588800 \pi ^8)\ , \\
  Z(17)&= \left(3171011760+23555952000 \pi -71723746080 \pi ^2-333199608000 \pi ^3+542885550648 \pi ^4+1355261623520 \pi ^5-\right. \\
      &\left.-1384280129304 \pi ^6-1337978574000 \pi ^7+518021476875 \pi ^8)\right/(3491169320514355200 \pi ^8) \ , \\
Z(18)&=(4970745360-180631896480 \pi ^2+2270514395520 \pi ^3+2444801550408 \pi ^4-\\
&-18251132155200 \pi ^5-13590443330584 \pi ^6+35949047139936 \pi ^7+\\
&\left.+20671882502409 \pi ^8-9607077219600 \pi ^9)\right/(377046286615550361600 \pi ^9) \ , \\
%
Z(19)&=(-2636096400+24895105200 \pi +79219113120 \pi ^2-487774106400 \pi ^3-\\
&-852843285000 \pi ^4+3053792290360 \pi ^5+3630439618136 \pi ^6-6122444513560 \pi ^7-\\
&\left.-4288974330849 \pi ^8+1840384320075 \pi ^9)\right/(55858709128229683200 \pi ^9)\ . 
\end{aligned}
\end{align}
}

Section \ref{sect2} of this paper is devoted to the derivation of this result.
Similar methods could be applied to $k>1$.
It would be interesting to find an analytic expression for 
general $k$ and $N$.

The knowledge of the exact values of $Z(N)$ in this paper allows one to
perform various numerical tests with high precision. As an example,
we compute in Section \ref{sect3} the coefficient of the membrane
1-instanton contribution to the partition function (see \eqref{inst}).

{\bf Note:} During the preparation of this manuscript
we received a paper \cite{Moriyama}, 
which has substantial overlap with our paper. 
The paper contains the exact results up to $N=9$,
which is consistent with ours.

\section{Derivation} \label{sect2}

Let us consider the grand canonical partition function 
\beq
\Xi(z)=1+\sum_{N\ge1} Z(N) z^N \ .
\label{Xidef}
\eeq
As is shown in \cite{Marino:2011eh},
this is given by a Fredholm determinant
\beq
\Xi(z)=\textrm{Det}\,\left(1+\frac{z \hat{K}}{4\pi}\right) \ ,
\eeq
with $\hat{K}$ defined by 
an integral kernel
\beq
K(x,y):=\langle x| \hat{K} | y \rangle=
\frac{e^{-u(x)-u(y)}}{\cosh\left( \frac{x-y}{2}\right)} \ ,
\eeq
and
\beq
u(x)=\frac{1}{2}\log \left( 2\cosh \frac{kx}{2}\right) \ .
\eeq
In practice, it is useful to use the following relation:
\beq
\Xi(z)=\exp\left( \textrm{Tr} \log \left(1+\frac{z
\hat{K}}{4\pi}\right)\right)=\exp\left(-\sum_\ell Z_\ell \frac{(-z)^\ell}{\ell}\right) \ ,
\label{XiZ}
\eeq
with
\beq
Z_\ell=\frac{1}{(4\pi)^\ell}\,\textrm{Tr}\,(\hat{K}^\ell) 
=\frac{1}{(4\pi)^\ell}\int dx_1 \ldots dx_\ell \, K(x_1, x_2) K(x_2, x_3) \ldots K(x_{\ell-1}, x_\ell) K(x_{\ell}, x_1)
\ .
\eeq
The problem thus reduces to the computation of $Z_\ell$.

Let us define the kernel for the operator $\hat{K} (I-\lambda^2 \hat{K}^2)^{-1}$
by $R_+(x, y)$ and for $\lambda\hat{K}^2 (I-\lambda^2 \hat{K}^2)^{-1}$ by
$R_-(x, y)$, respectively.
We also denote $R_+(x):=R_+(x,x), R_-(x):=R_-(x,x)$.
As is clear from the definition, the integral of $R_{\pm }(x)$ gives $Z_\ell$:
\beq
\frac{1}{4\pi}\int dx \, R_+(x)=\sum_{n\ge 0} (4\pi\lambda)^{2n}Z_{2n+1} \ , \quad
\frac{1}{4\pi}\int dx \, R_-(x)=\sum_{n\ge 0} (4\pi\lambda)^{2n+1}Z_{2n+2} \ . \label{R-Z-rel}
\eeq
Let us further define $\epsilon(\theta), \eta(\theta)$ by 
\beq
e^{-\epsilon(\theta)}=R_+(\theta)\  ,\quad
\eta(\theta)=2\lambda \int_{-\infty}^{\infty} \frac{e^{-\epsilon(\theta')}}{\cosh(\theta-\theta')}
d\theta'  \ . \label{eta-epsilon-rel}
\eeq
It was conjectured in \cite{Zamolodchikov:1994uw} and later proven in \cite{Tracy:1995ax} that these functions satisfy the 
following two TBA-like equations:
\begin{align}
&\epsilon(\theta)=2 u(\theta)-\frac{1}{2\pi} \int_{-\infty}^{\infty} 
\frac{\log\left(1+\eta^2(\theta')\right)}{\cosh(\theta-\theta')}
d\theta' \ ,\label{epsilon-eta-rel} \\
&R_-(\theta) =\frac{1}{\pi} R_+(\theta) \int_{-\infty}^{\infty} 
\frac{\arctan \eta(\theta')}{\cosh^2(\theta-\theta')} d\theta' \ .
\end{align}

Let us define
\begin{align}
&\epsilon(\theta)=\sum_{n\geq 0}\epsilon_n(\theta)\lambda^{2n}  \ , \quad
\eta(\theta)=\sum_{n\geq 0}\eta_n(\theta)\lambda^{2n+1} \ , \\
&R_+(\theta)=\sum_{n\geq 0}R_{+,n}(\theta)\lambda^{2n} \ , \quad
R_-(\theta)=\sum_{n\geq 0}R_{-,n}(\theta)\lambda^{2n+1} \ .
\end{align} 
Suppose $\epsilon_n(\theta),\,n=0\ldots j$ are known. We can then find $\eta_n(\theta),\,n=0\ldots j$ by performing the integration in (\ref{eta-epsilon-rel}), and then $\epsilon_{j+1}(\theta)$ from (\ref{epsilon-eta-rel}). Thus one can solve the TBA-like equations recursively, order by order in $\lambda$, starting from
\begin{equation}
\epsilon_0(\theta)=2u(\theta) \ .                                                                                                                                                                                                                                                                                                                                                                                                                                                                                                                                                                                                                                                                                         \end{equation} 
Once we know $\epsilon_n(\theta)$ and $\eta_n(\theta)$ for $n=0\ldots N$ we can find $R_{+,n}(\theta)$ and $R_{-,n}(\theta)$ for $n=0\ldots N$ and, therefore, 
$Z_{2n+1}$ and  $Z_{2n+2}$ for $n=0\ldots N$ from (\ref{R-Z-rel}).

Practically, it is useful to make the 
following change of variables: $e^\frac{\theta}{2}=t$. Then the equations (\ref{eta-epsilon-rel}-\ref{epsilon-eta-rel}) read
\begin{align}
&\eta(t)=8\lambda\int_0^\infty \! ds
 \,\,\frac{t^2s\,e^{-\epsilon(s)}}{s^4+t^4} \ , \label{epsilon} \\
&\epsilon(t)=\log\left(t^k+\frac{1}{t^k}\right)-\frac{2}{\pi}\int_0^\infty
 \! ds\,\,\frac{t^2s\,\log(1+\eta^2(s))}{s^4+t^4} \ . \label{eta}
\end{align}
Let us specialize to $k=1$ for simplicity. One can show that the functions $\epsilon_n(t)$, $\eta_n(t)$ have rather simple structure: 
\begin{align}
&\epsilon_n(t)=\sum_{j=0}^{n-1} G^{(n)}_j(t)(\log t)^j, \;n\geq 2 \ , \\
&\eta_n(t)=\sum_{j=0}^{n} H^{(n)}_j(t)(\log t)^j, \;n\geq 0 \ ,
\end{align} 
where $G^{(n)}_j(t)$ are rational functions with poles allowed at the roots of
$t^4-1$ and $H^{(n)}_j(t)$ are rational functions with poles allowed at the
roots of $t^4+1$. This observation allows easy calculation of the
integrals (\ref{epsilon}-\ref{eta}) order by order in $\lambda$ 
with the help of the residue theorem and the following formula:
\begin{equation}
 \int_0^\infty \!\! dt\, \, C(t)\, (\log t)^j=-\frac{(2\pi i)^j}{j+1}\int_\gamma\,  dt\,\, C(t)\, B_{j+1}\left(\frac{\log t}{2\pi i}\right) \ ,
\end{equation} 
where in the right hand side $\log t$ has a brunch cut from $0$ to $+\infty$, the contour $\gamma$ goes from $+\infty$ to $0$ below the cut and then to $+\infty$ above the cut, $C(t)$ is a rational function and $B_{j+1}(x)$ is Bernoulli polynomial.

Using the described procedure we find\footnote{This result agrees perfectly with the 
numerical result for $N\le 16$ in \cite{Moriyama}.}:
{\small
\begin{equation}
\begin{split}
Z_1&=\frac{1}{4} \ , \\
Z_2&=\frac{-2+\pi }{16 \pi }\ ,\\
Z_3&=\frac{\pi -3}{16 \pi }\ , \\
Z_4&=\frac{-4-8 \pi +3 \pi ^2}{128 \pi ^2}\ , \\
Z_5&=\frac{10-\pi ^2}{256 \pi ^2} \ , \\
Z_6&=\frac{36-2 \pi -3 \pi ^2}{1536 \pi ^2} \ , \\
Z_7&=\frac{-42+126
   \pi +49 \pi ^2-27 \pi ^3}{9216 \pi ^3}\ , \\
Z_8&=\frac{-96+96 \pi +64 \pi ^2-27 \pi ^3}{18432 \pi ^3}\ , \\
Z_9&=\frac{12-96 \pi -20 \pi ^2+5 \pi ^4}{32768
   \pi ^4}\ , \\
Z_{10}&=\frac{1200-2400 \pi -1400 \pi ^2+226 \pi ^3+135 \pi ^4}{1474560 \pi ^4}\ , \\
Z_{11}&=\frac{-660+23100 \pi -12100 \pi ^2-25300 \pi ^3-6303 \pi ^4+4725 \pi
   ^5}{29491200 \pi ^5}\ , \\
Z_{12}&=\frac{-720+3600 \pi +1200 \pi ^2-2560 \pi ^3-1536 \pi ^4+675 \pi ^5}{7372800 \pi ^5}\ , \\
Z_{13}&=\frac{4680-561600 \pi +978900 \pi ^2+655200 \pi ^3+10114 \pi
   ^4-30375 \pi ^6}{4246732800 \pi ^6}\ , \\
Z_{14}&=\frac{141120-1693440 \pi +764400 \pi ^2+1764000 \pi ^3+625436 \pi ^4-162882 \pi ^5-70875 \pi ^6}{14863564800 \pi ^6}\ , \\
Z_{15}&=\frac{-2520+1076040 \pi -4024860 \pi
   ^2-1425900 \pi ^3+2429714 \pi ^4+2860522 \pi ^5+527265 \pi ^6-509355 \pi
   ^7}{55490641920 \pi ^7}\ , \\
Z_{16}&=\frac{-40320+1128960 \pi -1599360 \pi ^2-1646400 \pi ^3-238336 \pi ^4+1136128 \pi ^5+663552 \pi ^6-297675 \pi ^7}{52022476800 \pi ^7} \ , \\
Z_{17}&=(85680-124750080 \pi +931227360 \pi ^2-303878400 \pi
   ^3-1054571336 \pi ^4-405544384 \pi ^5+\\
&\left.+45621608 \pi ^6+19348875 \pi^8)\right/(53271016243200 \pi ^8) \ , \\
Z_{18}&=(80640-5160960 \pi +15617280 \pi ^2+6397440 \pi ^3-10554208 \pi ^4-11079488 \pi ^5-\\
&\left.-2895216 \pi ^6+1060922 \pi ^7+385875 \pi ^8)\right/(1479750451200 \pi ^8) \ , \\
Z_{19}&=(-287280+1414279440 \pi -20169928800 \pi ^2+24409032480 \pi ^3+31396649256 \pi ^4+1177819272 \pi ^5-\\
&\left.-19209555560 \pi ^6-17783325576 \pi ^7-2533741371 \pi ^8+3094331625 \pi ^9)\right/(5753269754265600 \pi ^9) \ . \\
\end{split}
\end{equation}
}
From \eqref{Xidef}, \eqref{XiZ}
we obtain our main results \eqref{main}.

\section{Numerical Applications} \label{sect3}

It is easy to check numerically (cf. \cite{hanada,Moriyama}) that the
exact results obtained in the previous section are in agreement with the Airy function asymptotics. According to \cite{fhm,Marino:2011eh,hanada} the perturbative part of the partition function for $k=1$ is given by
\begin{figure}
\centering
\includegraphics{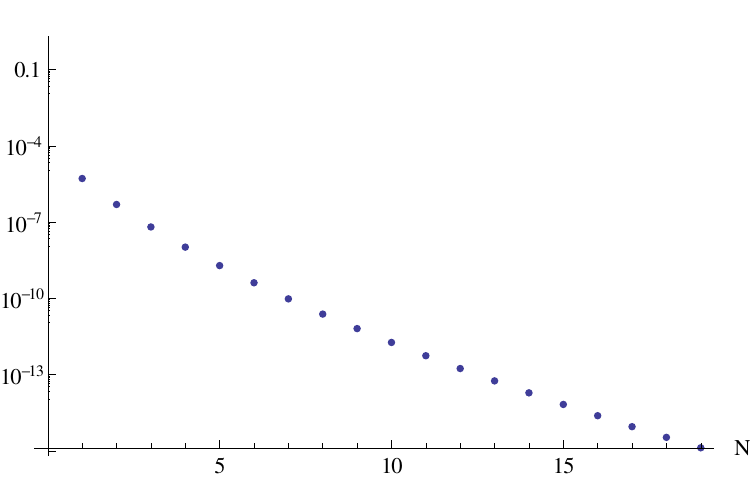}
\caption{In this figure, the dots represent the sequence $Z(N)/Z^\text{(pert)}(N)-1$.}
\label{figairytest}
\end{figure} 
\begin{equation}
 Z^\text{(pert)}(N)=C\,e^A\,\mathrm{Ai}\left[C\left(N-\frac{1}{3}-\frac{1}{24}\right)\right],
\end{equation} 
where 
\begin{equation}
 C=\frac{\pi ^{2/3}}{2^{1/3}},
\end{equation} 
\begin{equation}
 A=-\frac{\zeta(3)}{8\pi^2} + \frac{1}{6}\,\log\frac{\pi}{2} + 2\zeta'(-1) + \frac{1}{2}\log 2
-\frac{1}{3}\int_0^\infty dx\,\frac{1}{e^x-1}\left(\frac{3}{x^3}-\frac{1}{x}-\frac{3}{x\,(\sinh x)^2}\right).
\end{equation} 
The Fig.~\ref{figairytest} shows that indeed $Z(N)$ approaches $Z^\text{(pert)}(N)$ exponentially fast. In \cite{Moriyama} it was checked that the non-perturbative part $Z^\text{(np)}\equiv Z-Z^\text{(pert)}$ is suppressed by $e^{-2\pi\sqrt{2N}}$ which agrees with the previous analytical results \cite{dmp,Marino:2011eh,Drukker:2011zy}. One can go further and find the leading behavior of the prefactor. Namely, let us consider the following sequence:
\begin{equation}
 \tilde{F}(N)\equiv\left.\frac{Z^\text{(np)}(N)}{Z^\text{(pert)}(N)}\right/N\,e^{-2\pi\sqrt{2N}}=
\left.\left(\frac{Z(N)}{Z^\text{(pert)}(N)}-1\right)\right/N\,e^{-2\pi\sqrt{2N}}
\ .\label{nptest}
\end{equation} 
\begin{figure}
\centering
\includegraphics{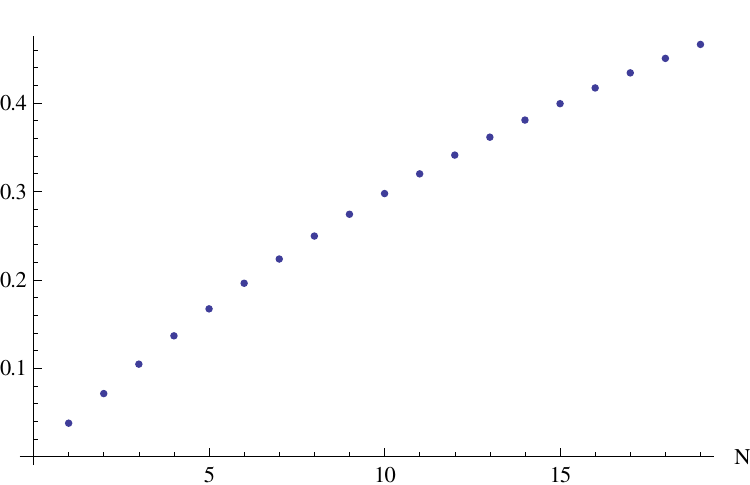}\;\includegraphics{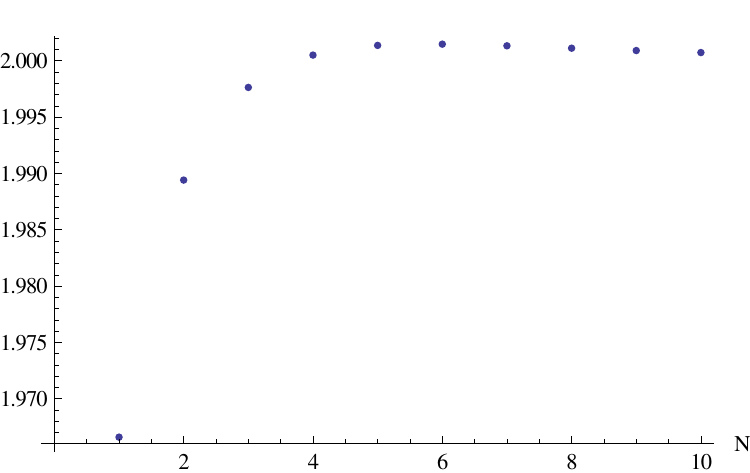}
\caption{In these figures, the dots represent the sequence (\ref{nptest}) (left) and its 9-th Richardson-like transform $\tilde{F}^{(9)}(N)$ (right).}
\label{fignptest}
\end{figure} 
From the previous works one expects that this sequence has an asymptotic expansion of the following form:
\begin{equation}
  \tilde{F}(N)=c_0+\frac{c_1}{N^{1/2}}+\frac{c_2}{N}+\frac{c_3}{N^{3/2}}+\ldots
 \ .\label{npteststr}
\end{equation} 
This assumption will be verified numerically \textit{a posteriori}. The
graph of $\tilde{F}(N)$ is shown on the left of
Fig.~\ref{fignptest}. One can accelerate convergence of the sequence
$\tilde{F}(N)$ by performing Richardson-like transforms. Let us define
the Richardson-like transform $\mathrm{R}_{\gamma}$ of a sequence $S(N)$ as
\begin{equation}
 \mathrm{R}_\gamma\left[S(N)\right]\equiv \left(N/\gamma+1\right)S(N+1)-NS(N)/\gamma \ . 
\end{equation} 
Its crucial property is that
\begin{equation}
 \mathrm{R}_\gamma\left[c+O\left(\frac{1}{N^\gamma}\right)\right]=c+o\left(\frac{1}{N^\gamma}\right).
\end{equation} 
In particular, if we define
\begin{equation}
  \tilde{F}^{(n)}(N)\equiv
   \mathrm{R}_\frac{n}{2}\left[\tilde{F}^{(n-1)}(N)\right] \quad (n\ge 1)
\ , \qquad
 \tilde{F}^{(0)}(N)\equiv \tilde{F}(N) \ ,
\end{equation} 
then one can show that 
\begin{equation}
 \tilde{F}^{(n)}(N)=c_0+O\left(\frac{1}{N^\frac{{n+1}}{2}}\right) \ .
\end{equation} 
The graph of $\tilde{F}^{(9)}(N)$ is shown on the right of
Fig.~\ref{fignptest}. The sequence converges very fast, which
verifies the self-consistency of the assumption (\ref{npteststr}).
Our numerical result suggests that $c_0=2$ exactly. One can also
numerically obtain $c_1,\,c_2,\ldots$ using similar techniques. On the M-theory side of the AdS/CFT correspondence this gives the 1-instanton contribution from M2-branes\footnote{The smallest instanton action is twice the action of the D2-brane considered in \cite{Drukker:2011zy} for $k=1$ because there is no M2 in $S^7$ which is a degree one cover of the D2-brane wrapping $\mathbb{RP}^3\in \mathbb{CP}^3$. However, there is an M2 wrapping $S^3\in S^7$ which is a degree two cover.}:
\begin{equation}
 \frac{Z^\text{(1-inst)}}{Z^\text{(pert)}}=\left(2N+O(\sqrt
					    N)\right)\,e^{-2\pi\sqrt{2N}}
 \ .\label{inst}
\end{equation} 
It would be interesting to check this by a direct calculation of
1-instanton contribution in M-theory. Let us note that the prefactor in
(\ref{inst}) cannot be obtained by previously developed techniques since 
they provide the non-perturbative part of the partition function as non-trivial asymptotic expansions either for large $k$ \cite{dmp} or for small $k$ \cite{Marino:2011eh}, whereas the result (\ref{inst}) is for $k=1$.


\section*{Acknowledgments}
The authors would like to thank Aspen Center for Physics (NSF Grant No.~1066293) for hospitality. P.~P. is supported by the Fonds National Suisse, subsidies 200020-126817 and 200020-137523, and by FASI RF 14.740.11.0347.



\begin{thebibliography}{99}
\bibliographystyle{plain}
\parskip=-2pt
 
    \bibitem{kwy}
A.~Kapustin, B.~Willett and I.~Yaakov, 
  JHEP {\bf 1003}, 089 (2010)
  [arXiv:0909.4559 [hep-th]].
  
\bibitem{jafferis} 
  D.~L.~Jafferis,
  JHEP {\bf 1205}, 159 (2012)
  [arXiv:1012.3210 [hep-th]].

\bibitem{hama}
N.~Hama, K.~Hosomichi, S.~Lee, 
  JHEP {\bf 1103}, 127 (2011)
  [arXiv:1012.3512 [hep-th]].
  
    \bibitem{kt}
I.~R.~Klebanov, A.~A.~Tseytlin, 
  Nucl.\ Phys.\  {\bf B475}, 164-178 (1996)
  [hep-th/9604089]. 

 \bibitem{mp}
  M.~Mari\~no and P.~Putrov, 
  JHEP {\bf 1006}, 011 (2010)
  [arXiv:0912.3074 [hep-th]].
 
 \bibitem{dmp}
 N.~Drukker, M.~Mari\~no, P.~Putrov, 
  Commun.\ Math.\ Phys.\  {\bf 306}, 511-563 (2011)
  [arXiv:1007.3837 [hep-th]]. 
  
 \bibitem{hklebanov}
 C.~P.~Herzog, I.~R.~Klebanov, S.~S.~Pufu, T.~Tesileanu, 
  Phys.\ Rev.\  {\bf D83}, 046001 (2011)
  [arXiv:1011.5487 [hep-th]].
  
  \bibitem{abjm}
  O.~Aharony, O.~Bergman, D.~L.~Jafferis and J.~Maldacena, 
  JHEP {\bf 0810}, 091 (2008)
  [arXiv:0806.1218 [hep-th]].

\bibitem{Drukker:2011zy} 
  N.~Drukker, M.~Mari\~no and P.~Putrov,
  JHEP {\bf 1111}, 141 (2011)
  [arXiv:1103.4844 [hep-th]].

\bibitem{Marino:2011eh}
  M.~Mari\~no and P.~Putrov,
  J.\ Stat.\ Mech.\  {\bf 1203} (2012) P03001
  [arXiv:1110.4066 [hep-th]].
  
\bibitem{Zamolodchikov:1994uw} 
  A.~B.~Zamolodchikov,
  Nucl.\ Phys.\ B {\bf 432}, 427 (1994)
  [hep-th/9409108].

\bibitem{Tracy:1995ax} 
  C.~A.~Tracy and H.~Widom,
  Commun.\ Math.\ Phys.\  {\bf 179}, 667 (1996)
  [solv-int/9509003].

\bibitem{okuyama}
  K.~Okuyama,
  Prog.\ Theor.\ Phys.\  {\bf 127}, 229 (2012)
  [arXiv:1110.3555 [hep-th]].
  
\bibitem{hanada}
  M.~Hanada, M.~Honda, Y.~Honma, J.~Nishimura, S.~Shiba and Y.~Yoshida,
  JHEP {\bf 1205}, 121 (2012)
  [arXiv:1202.5300 [hep-th]].

\bibitem{fhm}
H.~Fuji, S.~Hirano, S.~Moriyama, 
  JHEP {\bf 1108}, 001 (2011) 
  [arXiv:1106.4631 [hep-th]].

\bibitem{Moriyama} 
  Y.~Hatsuda, S.~Moriyama and K.~Okuyama,
  ``Exact Results on the ABJM Fermi Gas,''
  arXiv:1207.4283 [hep-th].


\end{thebibliography}
\end{document}